\documentstyle[aps,prl,epsfig]{revtex}
\begin{document}
\draft

\twocolumn[\hsize\textwidth\columnwidth\hsize\csname 
@twocolumnfalse\endcsname

\title{Effect of deconfinement on resonant transport in quantum 
wires}
\author{A. Ram\v{s}ak$^{1,2}$ and T. Rejec$^1$}
\address{$^1$J. Stefan Institute,  SI-1000 Ljubljana, Slovenia \\
$^2$Faculty of Mathematics and Physics, University of Ljubljana,
SI-1000 Ljubljana, Slovenia}
\author{J. H. Jefferson} 
\address{DERA,
Electronic Sector, St. Andrews Road, Great Malvern, Worcestershire WR14
3PS, England }

\date{\today}
\maketitle

\begin{abstract}
\widetext
\smallskip
The effect of deconfinement due to finite band offsets on transport
through quantum wires with two constrictions is investigated. It is
shown that the increase in resonance linewidth becomes increasingly
important as the size is reduced and ultimately places an upper limit
on the energy (temperature) scale for which resonances may be
observed. 
\end{abstract} 
 ]

\narrowtext
\section{Introduction}

Recent technological advances have enabled semiconductor
nanostructures to be fabricated with feature sizes down to tens of
angstroms. Such structures include arrays of `self-organized' quantum
dots and quantum wires. The former are grown by heteroepitaxial
deposition in which the over-layer material has the larger lattice
parameter and forms quantum dots to relieve elastic strain\cite
{tribe97,moison94,marzin94,grundmann95,grundmann95II,steer96,leonard}.
The latter may be achieved by 
heteroepitaxial growth on `v'-groove
surfaces, produced by optical lithography and etching, for which the
over-layer atoms diffuse preferentially towards the base of groove
producing a quantum wire with a crescent-shaped cross section
\cite{walther92,grundmann94,rinaldi94}. These structures have
potential opto-electronic applications, such as light-emitting diodes,
low-threshold lasers and single-electron devices.

In this paper we consider the ballistic transport of electrons through
quantum wires in which there are two constrictions defining a small
region of quasi-confinement. Such a system can have a rich and complex
resonance structure
\cite{nakazato91,kasai91,tekman93,gurvitz93,nokel94}. It
has been investigated in detail by Nakazato and
Blaikie\cite{nakazato91}, who consider characteristic sizes down to
$\sim $100~nm and assume that the quantum wire is defined by
infinitely high barriers. Such an assumption is reasonable provided
that the resonance levels considered are at sufficiently low-energy,
i.e. somewhat lower than that of the real band edge in wider-gap
material of the heterojunction. For the wire geometry they consider,
the resonances are very sharp at $T=0$. Inter-channel (mode) mixing of
the electron waves is important for higher-order resonances and can
give rise to unusual effects such as anti-resonances. At these
length-scales, however, the possible observation of such resonances
would be restricted to very low temperatures, where they would be
easily swamped by defects and disorder.

The main purpose of the present work is to investigate the resonance
structure for very narrow wires for which the energy (temperature)
scale is higher and may even approach room temperature. Such a
possibility is, of course, extremely important for device
applications. As the size shrinks, the effect of a finite barrier for
an electron in the wire becomes increasingly relevant, the main effect
being to broaden the resonances and restrict their number. Indeed, for
a sufficiently small confining region, only a single isolated
resonance level remains, with a continuum at higher energies. The
increasing width of the resonance level is due to deconfinement of the
electron wave-function into the classically forbidden region beyond the
potential step. However, all energies increase with increasing
confinement and the important criterion is whether the lowest
resonance peak can be resolved from the next resonance peak (or the
continuum), which sets the temperature scale. To be specific we shall,
in what follows, mainly consider two-dimensional structures with a
conduction-band offset of 0.4~eV, which is approximately the maximum
offset in the conduction band (direct gap) for GaAs wires on an
Al$_{x}$Ga$_{1-x}$ As.

\section{Model and method}

For simplicity, we restricted the model to two dimensions with
confinement in the $y$-direction and propagating in the $x$-direction.
The wire shapes under consideration are symmetric around the $x$ axis
and shown in Fig.~1. The width of the wire is parametrized as
$a(x)=a_0-a_1\sin^2 2\pi x/a_2$ for $0 \le x \le a_2$ and $a(x)\equiv
a_0$ otherwise. For comparison we choose two generic geometries. In
Fig.~1(a) the confined region is only weakly coupled to the `leads'
and we thus expect strong (narrow) resonances. This structure is
expected to behave in a similar way to a quantum dot which is
separated from its leads by tunnel barriers. On the other hand, for
the geometry of Fig.~1(b), the constrictions are much smaller
resembling a wire with weak thickness fluctuations.

We model further the wire as two regions of constant potentials, $V=0$
within the wire 'boundary', and confining potential $V=V_{0}>0$
outside the wire. The corresponding two-dimensional Schr\"{o}dinger
equation reads 
\begin{equation} -{\frac{\hbar ^{2}}{2m^{\ast
}}}\!\left( {\frac{\partial ^{2}}{\partial x^{2}
}}\!+\!{\frac{\partial ^{2}}{\partial y^{2}}}\right) \!\Psi
(x,y)+V\Psi (x,y)=E\Psi (x,y).  \label{schrodinger2D} 
\end{equation}
Here the electron effective mass $m^{\ast }$ is chosen to be that of
GaAs, equal to $0.067$ times the free electron mass and $E$ is
electron energy relative to the conduction band edge in the wire.

\begin{figure}[htb]
\begin{center}
\leavevmode\epsfxsize=90mm\epsfbox{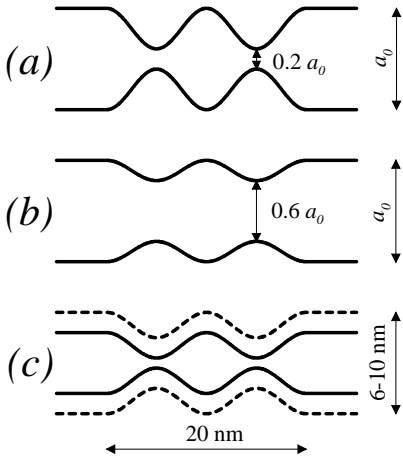}\end{center}
\caption{The width of the wire is always parametrized as
$a(x)=a_0-a_1\sin^2 2\protect\pi x/a_2$. (a) Weaker coupling: $a_1=0.8
a_0$ and $a_2=2 a_0$. (b) Stronger coupling: $a_1=0.4 a_0$ and $a_2=2
a_0$. (c) Shape of the `dot' fixed with $a_1=5$~nm, $a_2=20$~nm and
various wire thickness $a_0=6 - 10$ ~nm.} \end{figure}

The wave function is expanded in elementary modes or channels
\cite{szafer89}, 
\begin{equation} \Psi (x,y)=\sum_{n}\Phi
_{n}(x,y)\psi _{n}(x),  \label{expand} 
\end{equation} 
where the basis
functions $\Phi _{n}(x,y)$ are orthogonal solutions of one-dimensional
Schr\"{o}dinger equations in the $y$ direction for fixed $x$, with
eigenenergies $E_{n}(x)$.

Substituting the expansion Eq.~(\ref{expand}) into the
Schr\"{o}\-din\-ger equation Eq.~(\ref{schrodinger2D}) leads to a set
of ordinary linear differential equations for the channel $n$ wave
function $\psi _{n}(x)$, 
\begin{eqnarray} {\frac{d^2 \psi_n(x)}{d
x^2}}&+& k_{n}^2(x)\psi_n(x)=  \nonumber \\ &=&\sum_{m }\left[b_{n
m}(x){\frac{d \psi_m(x) }{d x}} +a_{n m}(x)\psi_m(x) \right], 
\label{set} 
\end{eqnarray} 
where $k_{n}^{2}(x)=2m^{\ast
}[E-E_{n}(x)]/\hbar ^{2}$ and 
\begin{eqnarray} a_{nm}(x) &=&-\int \Phi
_{n}(x,y)\frac{\partial ^{2}\Phi _{m}(x,y)}{\partial x^{2}}\;dy, 
\nonumber \\ b_{nm}(x) &=&-2\int \Phi _{n}(x,y)\frac{\partial \Phi
_{m}(x,y)}{\partial x} \;dy.  \label{ab} 
\end{eqnarray} 
For the
geometry considered here, there are no real bound states in the $x$
-direction; only scattering states are relevant.

Zero temperature conductance is calculated using the Landauer formula
\cite {landauer57,buttiker86}, 
\begin{equation} G=G_{0}\;{\cal T}(E), 
\label{landauer} 
\end{equation} 
where $G_{0}=2e^{2}/h$ and $E$ is here
the Fermi energy of the electrons in the leads. The transmission
probability, ${\cal T}(E)$, is the sum of transmission probabilities
for all channels, $n$, at energy $E$, i.e., 
\begin{equation} {\cal
T}(E)=\sum_{n}{\cal T}_{n}(E),  \label{1transmission} 
\end{equation}
where 
\begin{equation} {\cal T}_{n}(E)=\sum_{m}|{t}_{nm}(E)|^{2} 
\label{2transmission} 
\end{equation} 
and $t_{nm}(E)$ is transition
amplitude for scattering from channel $n$ to channel $m$.

At finite temperatures the conductance is calculated using a
generalized Landauer formula \cite{bagwell89} 
\begin{equation}
G=G_{0}\int_{-\infty }^{V_{0}}{\cal T}(\epsilon )\left[
-\frac{\partial f(\epsilon \!-\!E,T)}{\partial \epsilon }\right]
\;d\epsilon , \label{landauerT} 
\end{equation} 
where $f(\epsilon
,T)=[1+\exp (\epsilon /k_{{\rm B}}T)]^{-1}$ is the usual Fermi
function. This form is also based on the assumption that motion within
the wire is ballistic, the effect of temperature being merely to
change the energy distribution in the leads, thus allowing electrons
above and below the Fermi energy to contribute to the conductance. For
narrow wires, the tail of the Fermi distribution with energy $\epsilon
>V_{0}$ can be significant and the contribution of these electrons to
conductance will depend on the size and properties of the barrier
region. If this region is large, then electron motion will be
diffusive and may be described by an effective conductivity,
proportional to $\delta n$, the number of electrons in the Fermi tail.
The conductance due to these electrons will be further inhibited by
rough-surface scattering for mesa structures produced by etching. In
this paper we shall only consider the conductance due to electrons
within the wire by introducing an energy cut-off at $\epsilon =V_{0}$,
as shown in Eq.~(\ref{landauerT}). Thermally activated electrons in
the barrier region, which give rise to a series conductance, will be
considered in future work.

As pointed out in Ref.~\cite{nakazato91}, the Schr\"{o}dinger equation
Eq.~(\ref{schrodinger2D}) and the expression for conductance,
Eq.~(\ref{landauerT} ), are invariant under the scaling
transformation, 
\begin{eqnarray} x,y &\rightarrow &\lambda x,\lambda
y,  \nonumber \\ E,V &\rightarrow &\lambda ^{-2}E,\lambda ^{-2}V, 
\label{scaling} \\ T &\rightarrow &\lambda ^{-2}T.  \nonumber
\end{eqnarray}

To solve the system of differential equations, Eq.~(\ref{set}), we
first fix the number of channels, $N$, and then determine the
eigenfunctions $\Phi _{n}(x,y)$ with corresponding eigenenergies
$E_{n}(x)$ for $n\leq N$. $N$ must, of course, be sufficiently large
to ensure convergence. For narrow wire this poses a problem since
channels with energy above the barrier, $V_{0}$, form a
quasi-continuum and because of inter-channel coupling [c.f. terms at
the r.h.s. of Eq.~(\ref{set})], these high-energy channels can have a
significant effect on the eigenstates of electrons confined to the
wire. This may be understood in a perturbation theoretic sense. An
electron in state $\psi _{n}(x)$ with $E_{n}(x)<V_{0}$ may make a
virtual transition to a state $\psi _{m}(x)$ in the continuum
$(E_{m}>V_{0})$ and such transitions become very important for
$V_{0}-E_{n}$ small, i.e. for confined states close to the top of the
barrier. These excursions into the quasi-continuum enhance
deconfinement into the classically forbidden regions. This is
particularly important near narrow constrictions, such as the `necks'
in figure 1(a). The main effect of this, and the leakage of the base
states $ \psi _{n}$ into the barrier region, is to broaden resonances
compared to cases with $V_{0}=\infty $, considered in
Ref.~\cite{nakazato91}. To model the quasi-continuum we introduce
infinite barriers at $y=\pm L/2$. The basis functions, $\Phi
_{n}(x,y)$, $x$-fixed, are the simple standing waves. Some care is
needed in optimizing $N$ and $L$ to ensure convergence. $L$ must be
made sufficiently large to ensure that the effect of finite barriers
on the confined wave-functions is negligible. On the other hand, $L$
cannot be made too large otherwise the number of required channels
with $E>V_{0}$ becomes impractical. For the cases we considered,
$L\leq 20a_{0}$ and $N\leq 30$ were sufficient to ensure convergence.
The analytic expressions for the functions $\Phi _{n}(x,y)$ enable the
coefficients $a_{nm}(x)$ and $b_{nm}(x) $ in Eq.~(\ref{set}) to be
computed efficiently. The system of differential equations
Eq.~(\ref{set}) belongs to a class of `stiff' equations for which
direct integration is generally not stable. The basic problem is
exponentially increasing solutions with imaginary wave vectors of
different orders of magnitude leading to round off errors and
divergent results. This problem was solved for the present system by
dividing the wire along $x$-direction in $M$ sections. For each
section we first determined $2N $ independent solutions using the
fifth order Runge-Kutta numerical method. The length of each section
was chosen to ensure stable numerics in that section, the limiting
factor being the number of channels with imaginary $k_{n}(x)$. In
our case up to $10$ such channels were taken into account within each
section, with $M\leq 10$ sections. Matching the solutions at each
boundary yields sets of linear equations from which the transition
amplitudes, $t_{nm}(E),$ may be determined.

\section{Results}

Fig.~2(a) shows electron conductance at $T=0$ versus energy for the
dot geometry of Fig.~1(a), i.e., $a_{1}=0.8a_{0}$ and $a_{2}=2a_{0}$
for wire widths $a_{0}$ from 4~nm to 20~nm and the barrier height
$V_{0}=0.4$~eV. The open circles, $\circ $, in this and the remaining
figures correspond to a Fermi energy $E=V_{0}$, as explained in the
previous section, electrons with energy greater than this are not
'bound' to the wire and their conductance will be dominated by the
properties and size of the barrier region. For convenience of
presentation and to emphasize the effects of scaled units of energy,
we choose scaled units of energy, $E/E_{0}$, where $E_{0}=\hbar
^{2}/(2m^{\ast }a_{0}^{2})$, the ground-state energy for an electron
in a one-dimensional well of width $a_{0}$ with infinitely high
potential walls. For perfect confinement $(V_{0}=\infty )$, the
scaling invariance, Eq.~(\ref {scaling}), at $T=0$ gives,
\begin{equation} G(\lambda a_{0},E/E_{0}(\lambda
a_{0}))=G(a_{0},E/E_{0}(a_{0})) \end{equation} and hence all wires of
the same shape have identical conductance curves for $V_{0}=\infty
$\cite{nakazato91}. This universal curve is approximately that for
$a_{0}=20$~nm in Fig.~2(a) (for which the finite barrier height is
irrelevant). We note that the (first) resonance occurs at $E\sim
2.5E_{0}$, which may be interpreted approximately as a resonant bound
state with energy $E_{0}$ due to confinement in the $y$-direction and
energy $1.5E_{0}$ due to quasi-confinement in the x-direction.

We see that the effect of finite $V_{0}$ is to shift the position of
the first resonance to lower energies with decreasing wire width. This
is, of course, due directly to the deconfinement effect of a finite
band offset. For very narrow wires the resonant level is pushed
towards the continuum (i.e. unbound in 2D) at energy $E=V_{0}=0.4$~eV
and only one resonant level is possible. This is seen to be the case
for wires of width 4, 5 and 7~nm. The corresponding energies from the
peak to the continuum are 83, 151 and 232~meV for these wires. For
wider wires the energy splitting between the first two peaks reduce in
energy, eventually decreasing like $a_{0}^{-2}$ (approximately).
Furthermore, the resonances always become broader with decreasing wire
width, again a consequence of the deconfinement effect of finite band
offset. Thus, at zero temperature, multiple resonances with the
highest resolution (ratio of resonance separation to resonance width)
occur for wide wires, the limiting resonance width being determined by
the geometry (strength of the effective tunnel barriers into the
confined region). However, the absolute energy scale is, of course,
very small and these sharp resonances are rapidly broadened with
temperature. This is shown in Fig.~2(b) where for the same wires we
plot conductance at $T=100$~K and 300~K with thick and thin lines,
respectively. Open circles $\circ $ again represent an energy cut-off
$E=V_{0}$, above which conduction is primarily through the barrier
region. The thermal broadening is accompanied by a reduction in the
peak heights which are barely resolvable at room temperature for wires
with $a_{0}>10$~nm, though clearly resolvable and optimum for
$a_{0}\sim 5$~nm. This is significantly better than the 4~nm width
wire for which the proximity of the continuum has a large effect.

\begin{figure}[tbh]
\begin{center}
\leavevmode\epsfxsize=80mm\epsfbox{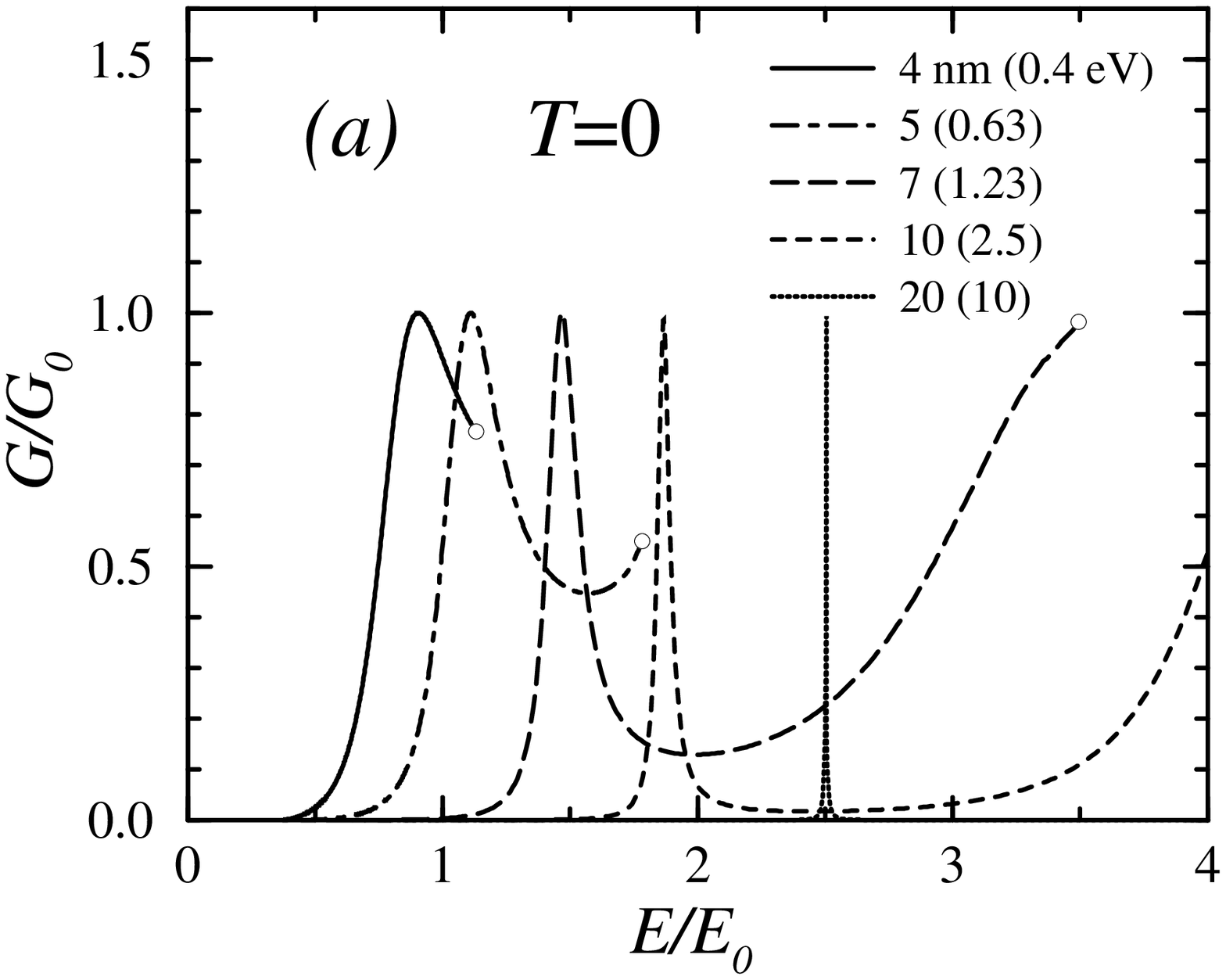}\end{center}
\begin{center}
\leavevmode\epsfxsize=80mm\epsfbox{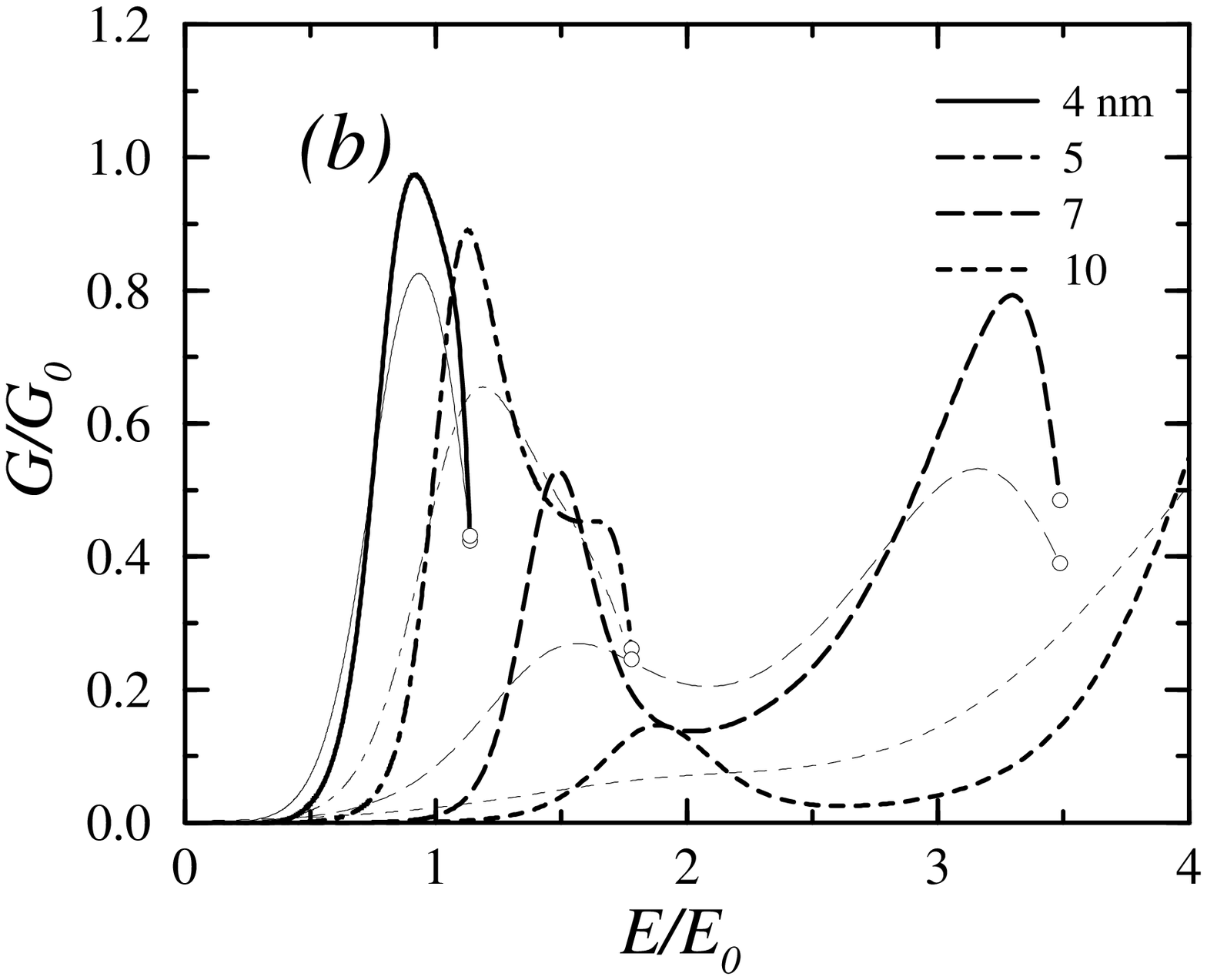}\end{center}
\caption{(a) $T=0$ conductivity [geometry from Fig.~1(a)] for various
$ a_{0}=4-20$~nm and fixed $V_{0}=0.4$~eV (or, equivalently, with
fixed $ a_{0}=4$~nm and various $V_{0}=0.4-10$~eV). (b) Conductivity
at $T=100$~K and 300~K presented with thick and thin lines,
respectively, for fixed $V_{0}=0.4$~eV and various $a_{0}=4-10$~nm. 
Circles $\circ $ represent upper limit $E=V_{0}$ of
calculation.} \end{figure}

Because of the invariance under scaling, Eqs.~(\ref{scaling}), the
same behavior occurs for larger wires with a smaller band offset,
though the overall energy scale is lower. Thus Fig.~2(a) may be
regarded as a plot of conductance for wires with {\it same} width
(4~nm in this case) but {\it  different} band offsets, $V_{0}$.
Increasing $V_{0}$ increases the confinement and has the same effect
as increasing the wire width with fixed $V_{0}$ and visa versa,
apart from an overall change in energy/temperature scale. Indeed, the
effects of deconfinement could be investigated experimentally for
relatively large wires by fabricating quasi-2D wires with small
conduction-band offsets. The quasi-2D behavior would be achieved by
ensuring high confinement in the third dimension. The behavior of such
wires at low-temperatures, i.e. conductance versus energy/gate voltage
for various widths, would be the same as that of narrow wires at
higher temperatures.
\footnote{ This assumes very clean wires and
parabolic bands. In practice one would have to take into account
non-parabolicity effects for very narrow wires and the effects of
disorder would become increasingly important for larger wires. }

In Fig.~3 we show conductance plots at absolute zero and $T=100$~K for
the more weakly confining wire of Fig.~1(b), for which
$a_{1}=0.4a_{0}$ and $a_{2}=2a_{0}$. The behavior is seen to be
qualitatively similar to the more strongly confined geometry wire
though the resonances are broader, reflecting the weaker confinement
along the wire (smaller effective tunnelling barriers). However the
thermal broadening is still largely governed by the overall wire
thickness. For example, if we compare the 5~nm wires in Fig.~2(b) and
Fig.~3 at $T=100$~K (thin lines) we see a similar relative increase in
the halfwidth and decrease in the peak height compared with absolute
zero. This shows that even for small thickness fluctuations, resonance
peaks can persist to quite high temperatures for narrow wires.

\begin{figure}[htb]
\begin{center}
\leavevmode\epsfxsize=80mm\epsfbox{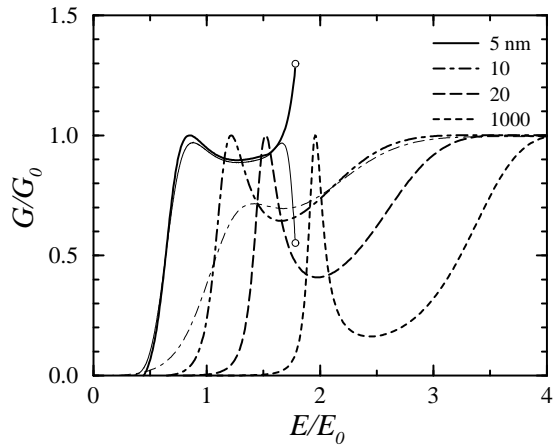}\end{center}
\caption{$T=0$ conductivity [geometry from Fig.~1(b)] for various
$a_0=5 - 1000$~nm and fixed $V_0=0.4$~eV is presented with thick
lines. Thin lines represent the corresponding $T=100$~K result.
Circles $\circ$ represent upper limit $E=V_0$ of calculation.}
\end{figure}

Finally we consider the effect of reducing only the thickness of a
quantum wire, $a_{0}$, whilst otherwise maintaining the same shape,
$a_{1}=5$~nm and $a_{2}=20$~nm, as shown in Fig.~1(c). This has the
effect of producing a quasi-1D quantum dot. There are two competing
effects as the wires is made narrower: the effective tunnel \hfil
\begin{figure}[htb]
\begin{center}
\leavevmode\epsfxsize=80mm\epsfbox{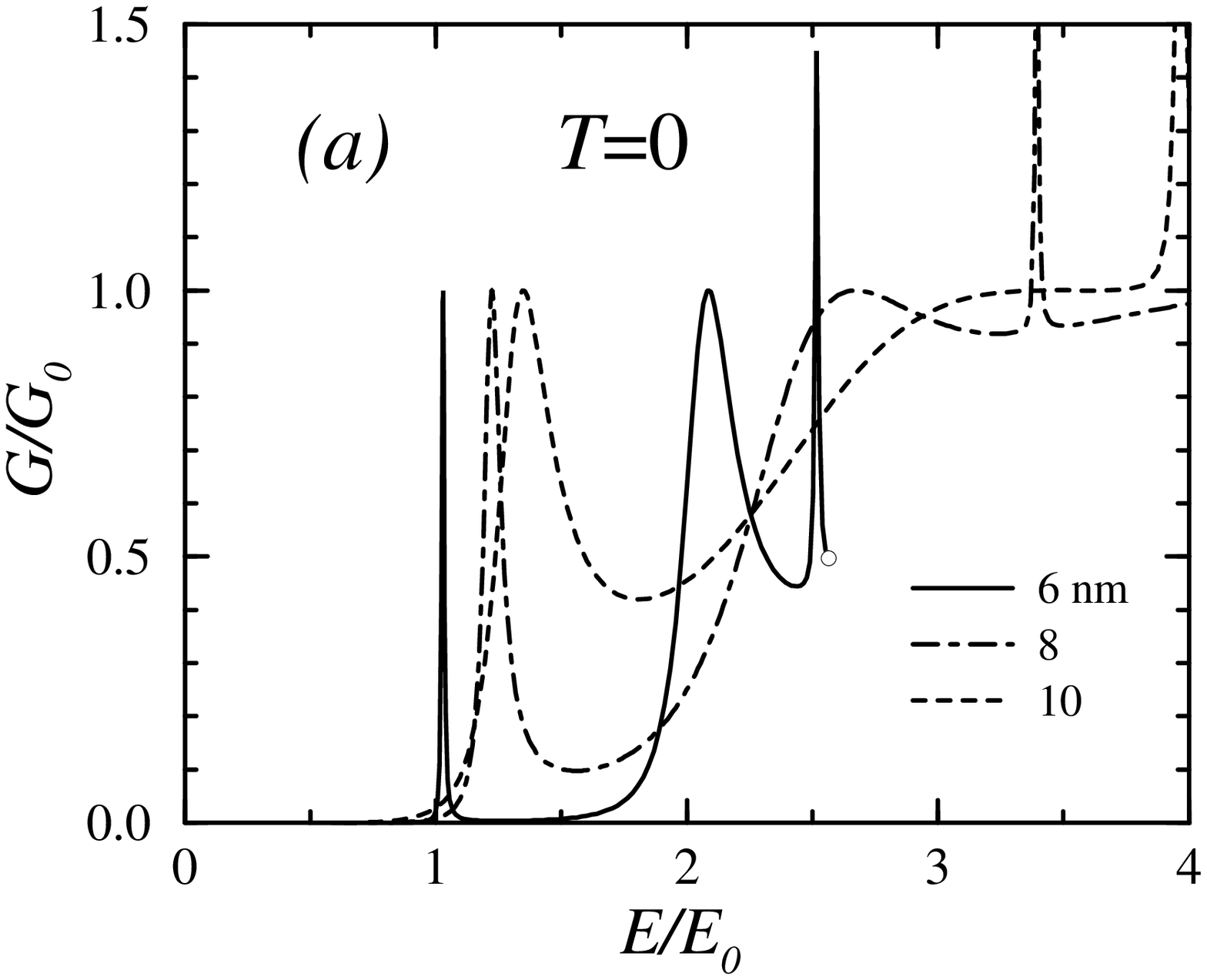}\end{center}
\begin{center}
\leavevmode\epsfxsize=80mm\epsfbox{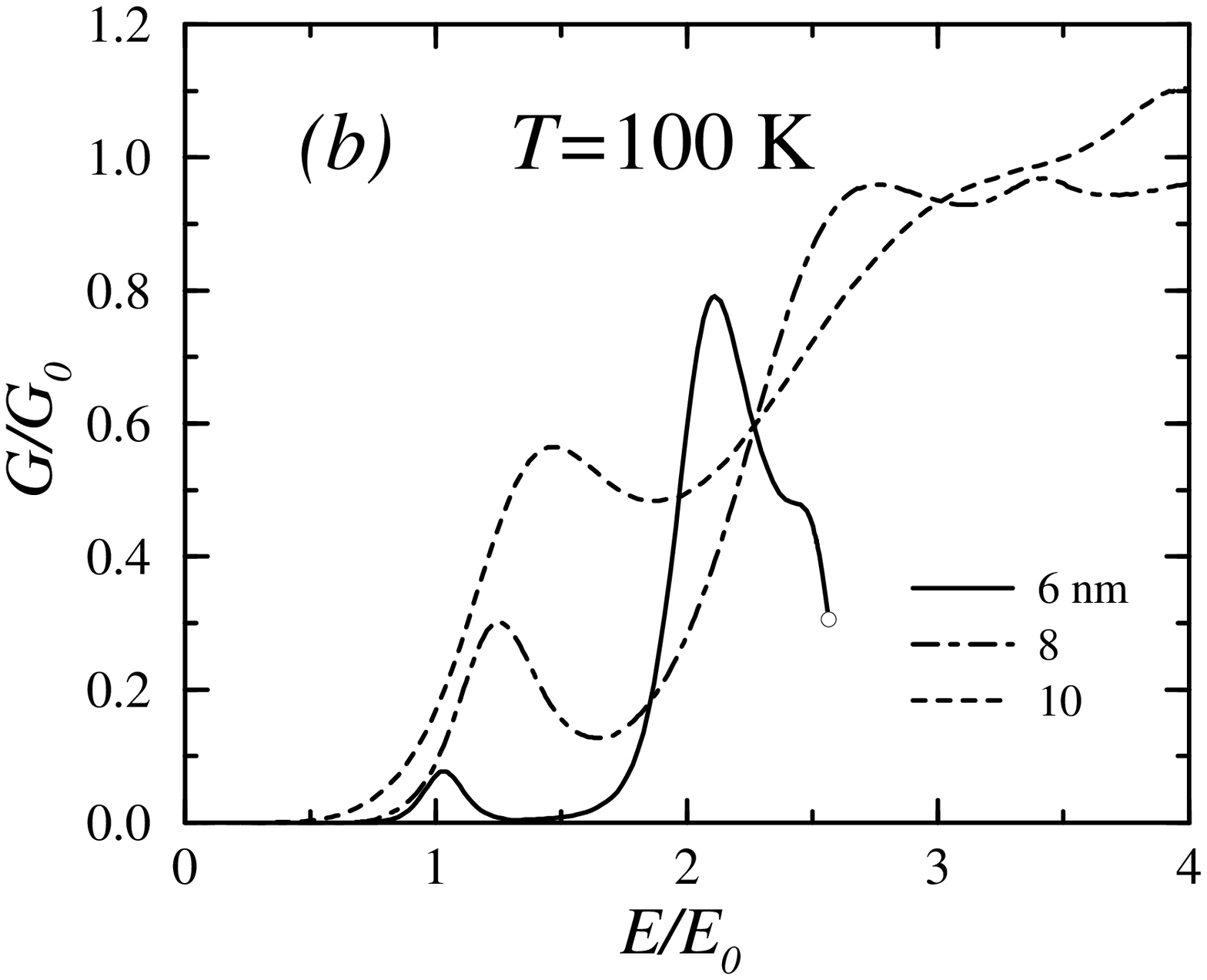}\end{center}
\caption{(a) $T=0$ conductivity [geometry from Fig.~1(c)] for various
$a_0=6 - 10$~nm and fixed $V_0=0.4$~eV. (b) Conductivity at $T=100$~K
and the same geometry. Circles $\circ$ represent upper limit $E=V_0$
of calculation.} \end{figure}
\noindent
barriers
increase and there is enhanced deconfinement near the necks. It turns
out that the increase in effective tunnel barriers is the greater
effect, resulting in sharper resonances as the wire thickness is
reduced, as shown in Fig.~4(a). This should be contrasted with
Fig.~2(a), which always gives rise to a resonance broadening when the
overall size is reduced. However, we point out that, unlike the case
of thick wires, this reduction in linewidth is significantly less than
that given by a single-channel approximation. This is because there is
coupling to the 2D continuum near the neck region. This coupling
depends on both the rate of narrowing of the wire and its absolute
width. The other main effect shown in Fig.~4(a) is the appearance of
further resonance peaks below the continuum. This reflects the
quasi-1D nature of the confinement region, the higher-lying peaks
corresponding to higher harmonics along the length of the effective
quantum dot. Their separation is set by the length of the confinement
region along the wire. Conductivity at finite a temperature $T=100$~K
is plotted in Fig.~4(b). We see that for the wire thicknesses
considered, the higher harmonics broaden rapidly with increasing
temperature, merging into the quasi-2D continuum. However, it can be
seen that the lowest resonance remains distinct.

\section{Conclusions}

With realistic conduction band offsets, the lowest resonance peak for
ballistic transport through quantum wires of fluctuating thickness
giving rise to a quasi quantum dot should be discernible around room
temperature for smallest structures close to the limits of present
fabrication techniques. Nevertheless, deconfinement effects due to
finite band offsets are significant and ultimately the limiting factor
for sufficiently small structures. In principle the resonances could
be made to survive to higher temperatures if heterojunctions could be
fabricated with even larger band offsets. The restriction to 2D in the
simulations is somewhat artificial here though the extension to true
3D with circular cross-section wires is feasible and calculations are
in progress to estimate the expected enhanced deconfinement effect
which they would produce. However, the general behavior is expected to
be similar to that described in this paper. Other geometries, such as
might be produced from a self-organized quantum dot connected to
source and drain contacts `vertically' is also expected to behave in a
similar fashion, though there would be quantitative differences of
course. Future work will consider other effects which become
increasingly important for very small structures, including
non-parabolicity and Coulomb blockade, particularly the effect of
deconfinement on the latter.

\acknowledgments
We acknowledge valuable discussions with Colin Lambert. This work was
partly funded by the EU (Contracts ERB CIBD CT940017 and
CHRX-CT93-1036) One of the authors (A. R.) would like to thank DERA
for the hospitality which was extended to him during his several
visits.

\end{document}